\documentclass[Times,twocolumn,tighten,deluxetable]{aastex631}

\usepackage{graphicx}
\usepackage{multirow}
\usepackage{txfonts}
\usepackage{natbib}
\usepackage{url}
\usepackage{hyperref}
\usepackage{longtable}
\usepackage{tabularx}
\usepackage{lipsum}
\usepackage{array}
\usepackage{rotating}
\usepackage{changepage}

\newcommand{\jetcaf}{{\fontfamily{qcr}\selectfont JeTCAF}}
\newcommand{\kerrbb}{{\fontfamily{qcr}\selectfont KERRBB}}
\newcommand{\dbb}{{\fontfamily{qcr}\selectfont DISKBB}}
\newcommand{\tcaf}{{\fontfamily{qcr}\selectfont TCAF}}

\newcommand{\tbabs}{{\fontfamily{qcr}\selectfont TBABS}}

\newcommand{\polcon}{{\fontfamily{qcr}\selectfont POLCONST}}
\newcommand{\constant}{{\fontfamily{qcr}\selectfont CONST}}
\newcommand{\plaw}{{\fontfamily{qcr}\selectfont POWERLAW}}
\newcommand{\cutoffpl}{{\fontfamily{qcr}\selectfont CUTOFFPL}}
\newcommand{\gauss}{{\fontfamily{qcr}\selectfont GAUSS}}

\shorttitle{Spectro-polarimetric study of Swift J151857.0-572147}
\shortauthors{Mondal et al.}

\begin{document}

\title{The first detection of X-ray polarization in a newly discovered Galactic transient Swift\,J151857.0-572147}

\correspondingauthor{Santanu Mondal}

\email{santanuicsp@gmail.com}

\author[0000-0003-0793-6066]{Santanu Mondal}
\affiliation{Indian Institute of Astrophysics, II Block, Koramangala, Bengaluru 560034, Karnataka, India}

\author{S. Pujitha Suribhatla}
\affiliation{Indian Institute of Astrophysics, II Block, Koramangala, Bengaluru 560034, Karnataka, India}

\author[0000-0002-6252-3750]{Kaushik Chatterjee}
\affiliation{South-Western Institue For Astronomy Research, Yunnan University, University Town, Chenggong, Kunming 650500, China}

\author[0000-0002-7782-5719]{Chandra B. Singh}
\affiliation{South-Western Institue For Astronomy Research, Yunnan University, University Town, Chenggong, Kunming 650500, China}

\author{Rwitika Chatterjee}
\affiliation{Space Astronomy Group, ISITE Campus, U. R. Rao Satellite Center, ISRO, Bengaluru, 560037, India}


\begin{abstract}
We study the spectro-polarimetric properties of a newly discovered black hole X-ray binary Swift\,J151857.0-572147 jointly using {\it IXPE} and {\it NuSTAR} observations during March 2024. The analysis of {\it IXPE} data reports the first detection of X-ray polarization with degree (PD) $1.34\pm0.27$ and polarization angle (PA) $-13.69^\circ\pm5.85^\circ$ using model-independent approach, while the model-dependent analysis gives PD $1.18\pm0.23$ and PA $-14.01^\circ\pm5.80^\circ$. The joint spectral analysis of the broadband data and {\it NuSTAR} analysis in isolation constrain the mass of the central black hole between $\sim 9.2\pm1.6-10.1\pm1.7 M_\odot$ and a moderate spin parameter of $\sim0.6\pm0.1-0.7\pm0.2$ with disk inclination $\sim 35^\circ\pm7^\circ-46^\circ\pm15^\circ$. The power-law photon index and cutoff energy are $2.19\pm0.03-2.47\pm0.06$ and $\sim 36\pm4-78\pm10$ keV, suggesting a transition to the soft spectral state (SS). Additionally, a relatively lower corona size of $6\pm1-9\pm2$ $r_S$, a low mass outflow rate ($<3$\% $\dot M_{\rm Edd}$), and the best-fitted halo accretion is less compared to the disk accretion rate further confirms the same state. The low PD detected in the SS can be due to repeated scattering inside the dense corona and the dominant emission from the disk, agrees with the low spin and low disk inclination. The hydrogen column density obtained from the fit is relatively high $\sim 4-5\times 10^{22}$ cm$^{-2}$.

\end{abstract}




\section{Introduction}\label{introduction}
Matter accreting onto a black hole (BH) forms an accretion disk consisting mainly of two components, the inner puffed-up region which is known as the so-called corona, and the cold Keplerian disk. The first component upscatters soft photons from the Keplerian disk through inverse Comptonization making them hard \citep[][and references therein]{SunyaevTitar80,HaardtMaraschi1993,ChakrabartiTitarchuk1995,Done2007,MondalChak2013MNRAS.431.2716M}. The same region also launches jet/outflows \citep{Chakrabarti1999,ChattopadhyayDas2007NewA...12..454C,MondalChakrabarti2021} and polarizes the radiation depending on its temperature, optical depth, geometry, and inclination \citep[][and references therein]{ConnorsEtal1980ApJ...235..224C,SunyaevTitar85}. In addition to the above geometrical configuration and accretion properties, the spin of the BH is a crucial parameter in deciding the amount of polarization as well as its energy dependence. Therefore, the polarization properties such as polarization angle (PA) and degree (PD) of any source depend on several factors including accretion properties and the spin of the BH. The variation of PD and PA on the optical depth, disk inclination, spin of the BH, and outflows was studied in detail both in theory and simulation \citep[][and references therein]{BegelmaMcKee1983ApJ...271...89B,DovciakEtal2008MNRAS.391...32D,LiEtal2009ApJ...691..847L,SchnittmanKrolik2010ApJ...712..908S, Krawczynski2012ApJ...754..133K,TavernaEtal2020MNRAS.493.4960T,Ratheesh2024ApJ...964...77R}. Therefore, by constraining the coronal properties of a system, the PD can be constrained or vice-versa along with the intrinsic parameters of the BH, making spectro-polarimetric studies important in recent time.

Black Hole X-ray binaries (BHXRBs) show variabilities in their spectral states during the outburst time \citep{RemillardMcClin2006ARA&A..44...49R}. In general, they follow a cycle starting from a hard state (HS) when spectra are dominated by hard photons from the corona region to a soft state (SS) when spectra are dominated by soft photons mainly from the disk through intermediate states \citep[][and references therein]{HomanEtal2001ApJS..132..377H,BelloniEtal2005A&A...440..207B,RemillardMcClin2006ARA&A..44...49R,NandiEtal2012A&A...542A..56N,DebnathElal2015MNRAS.447.1984D,ShuiEtal2021MNRAS.508..287S,Janaetal2016}.  Sometimes, some of the sources may not follow the cycle for several reasons e.g. lack of supply of matter or lack of viscosity etc. \citep{King1998MNRAS.296L..45K,ChakEtal2019AdSpR..63.3749C,Mondal2020AdSpR..65..693M,Mondaletal2017}. Outbursts that do not go to the SS \citep[][and references therein]{TetarenkoWATCHDOG2016ApJS..222...15T} or do not pass through all states are called failed-outburst. In the SS, the disk moves much closer to the BH and the emission from the corona region is negligibly small. As the corona nearly disappears in the SS and its optical depth increases, it can trap more radiation, lowering the probability of hard emission \citep[see][and references therein]{SunyaevTitar80,Chakrabarti1997,MondalChak2013MNRAS.431.2716M}. Therefore, the scattering can also affect the amount of polarization \citep{SunyaevTitar85}.

The new Galactic transient Swift J151857.0-572147 (hereafter SwiftJ151857) was first detected by Swift/XRT as a GRB \citep[GRB 20240303A;][]{KenneaEtal2024ATel16500....1K} in Swift Trigger 1218452 
\footnote{https://gcn.nasa.gov/circulars/35849}{(GCN 35849)}. However, its constant bright nature without signs of fading along with its localization in the Galactic plane was later confirmed to be a Galactic transient. From the best localization of the source using XRT immediate on-board localization, the RA, and Dec of the source were found to be RA(J2000) = 15h 18m 57.00s and Dec(J2000) = -57d 21` 47.9`` \citep{KenneaEtalB2024GCN.35853....1K}. Follow-up radio observations with the MeerKAT telescope were done \citep{CowieEtal2024ATel16503....1C} at $1.28$ GHz (L-band) with a bandwidth of $856$ MHz at a flux density of 10 mJy \citep{CarotenutoEtal2024ATel16518....1C} on 04 March 2024 for 15 minutes. The inverted radio spectrum ($f(\nu) \propto \nu^\alpha$, where $\nu  \sim +0.5$) with the photon index helped designate the nature of the source as consistent with that of an X-ray binary in the hard state, suggesting it may be a neutron star or a black hole. The Australia Telescope Compact Array (ATCA) followed-up the radio observations on 09 March 2024 for $\sim 30$ minutes at frequencies $5.5$ and $9$ GHz simultaneously \citep{SaikiaEtal2024ATel16516....1S}. Their analysis also affirmed that the source was a Galactic black hole. After the ATCA, Swift/XRT performed target of opportunity (ToO) on this source with an exposure of 1000s. It was found that the spectrum is well described by the combination of phenomenological disk black body (\dbb) and power-law model (\plaw) models \citep{DelSantoEtal2024ATel16519....1D}. These findings also solidified the nature of the source as a black hole. There was INTEGRAL serendipitous detection of the source from 8 to 11 March 2024 \citep{SgueraEtal2024ATel16524....1S}. As part of the monitoring program of GRBs in optical and NIR wavelengths, the 60 cm Robotic Eye Mount (REM) telescope monitored the source 
in both these wavelengths \citep{BaglioEtal2024ATel16506....1B}.

The broadband X-ray data of the source from {\it Insight-HXMT} \citep{ZhangEtalInsight2020SCPMA..6349502Z} was analyzed by \citet{ChatterjeEtal2024arXiv240617629C} in the intermediate spectral state. Authors detected the presence of type-C quasi-periodic oscillations (QPOs) during the outburst period, which disappeared in the SS while {\it The Imaging X-ray Polarimetry Explorer} \citep[IXPE;][]{WeisskopfEtal2016SPIE.9905E..17W} observed the source. The Swift/XRT spectral modeling of the source reported a column density ($N_H$) of $5.6\pm0.06\times10^{22}$ cm$^{-2}$ with a power-law photon index ($\Gamma$) of $1.78\pm0.02$ \citep[see][]{KenneaEtal2024ATel16500....1K}. A similar $N_H$ was estimated by \citet{ChatterjeEtal2024arXiv240617629C}. The negative velocity of an HI absorption line imposes the lower limit on the distance to the source to be $4.48^{+0.67}_{-0.47}$ kpc. While the absence of positive velocity absorption lines towards other sources in the field of the HI absorption for Swift\,J151857 puts an upper limit on the distance as $15.64^{+0.77}_{-0.60}$ kpc \citep{BurridgeEtal2024ATel16538....1B}. Due to a large uncertainty in distance estimation, we used an average distance of 10 kpc throughout our analysis for this source. Other information like the mass, inclination, and spin have not been reported about the source yet. From our analysis, we report these intrinsic properties of the source for the first time.

Recently, {\it IXPE} has observed several black hole X-ray binaries (BHXRBs) along with SwiftJ151857 and showcased their X-ray polarization detection. In 4U\,1630–47 PD $\sim6.5-10$\% was observed \citep{Rodriguez_Cavero_2023,KushwahaEtal2023MNRAS.524L..15K,Rawat2023MNRAS.525..661R,Ratheesh2024ApJ...964...77R}, in the HS of Swift\,J1727.8–1613 the PD measured $\sim3-4$\%   \citep{VeledinaEtal2023ApJ...958L..16V,IngramEtal2023MNRAS.525.5437I,PodgornyEtalSwift17272024A&A...686L..12P}, while a rapid decrease in PD $\sim0.5$\% was observed by \citet{SvobodaEtalSwiftJ17272024ApJ...966L..35S} in the SS. In LMC X-3, $\sim3$\% polarization was detected \citep{SvobodaEtalLMCX32024ApJ...960....3S,MajumderEtal2024MNRAS.527L..76M} and in the SS of 4U\,1957+115 measured PD was $\sim1.95$\% \citep{Marra2023arXiv231011125M,KushwahaEtal4U19572023arXiv231002222K}. A 4\% PD was detected in Cyg X-1 in the HS and $2.5-2.8$\% in the SS \citep{KrawczynskiEtal2022Sci...378..650K,JanaIXPE2024MNRAS.52710837J}. Very recently, \citet{VeledinaEtalCygX32023arXiv230301174V} detected a high polarization $\approx 25\%$ in Cyg X-3 using {\it IXPE} observation. These studies not only detected polarization but also showed the variation in PD in different spectral states, as expected from the simulations. Therefore, there has been renewed interest in studying BHXRBs to further understand their accretion geometry, emission mechanisms, and intrinsic parameters of the central BH. The SwiftJ151857 source was observed by {\it NuSTAR} \citep{Harrisonetal2013} simultaneously with {\it IXPE}. 

Motivated by recent X-ray polarization detections, which have been used as a powerful tool to understand several emission and accretion properties of BHXRBs, in this paper, we performed spectro-polarimetric studies of SwiftJ151857. We used both {\it IXPE} and {\it NuSTAR} observations of the source to carry out broadband spectral and polarimetric studies to understand the origin of X-ray polarization, its properties, spectral states, accretion geometry, and intrinsic parameters of the BH. Along with the corona properties, we have also estimated the mass outflow from this system, that connects the disk-corona-outflow with polarization in a single framework.    
The paper is organized as follows: in \S 2 we discuss the observation and data reduction procedure. In \S 3, we discuss the spectro-polarimetric
data analysis results, and the origin of the detected polarization, and finally, we draw our conclusions in \S 4.

\section{Observation and Data Reduction}
The source Swift\,J151857 was observed on 18 March 2024 by {\it IXPE} with exposure time 96 ks and {\it NuSTAR} observed on 17 and 18 March (hereafter, epochs N1 and N2) with exposures 14 and 9 ks respectively. We present the analysis of the data from both the days in this paper. The corresponding observation IDs are 03250201 (as epoch I), 91001311002 (as epoch N1), and 91001311004 (as epoch N2). 

{\it IXPE} is an imaging X-ray telescope consisting of 3 polarization-sensitive detector units (DUs) that observe in the 2-8 keV energy range. We obtained the Level-2 event files from the HEASARC archive \footnote{https://heasarc.gsfc.nasa.gov/docs/ixpe/archive/} which are cleaned and calibrated using standard \texttt{FTOOLS} tasks with the latest calibration files (CALDB 20230526) available in {\it IXPE} database. For extracting the source region, we select a circular region of $50^{\prime\prime}$ for all detectors.  

For {\it NuSTAR} analysis, we considered the observations from 17 and 18 March 2024 and have used the standard 
{\sc NUSTARDAS v1.3.1}
\footnote{https://heasarc.gsfc.nasa.gov/docs/nustar/analysis/} software to extract the data. We generated cleaned event files
using the {\sc nupipeline} task and generated the spectra using
{\sc nuproducts} along with CALDB version 20220331. We considered a region of 
$60^{\prime\prime}$  for the source and $120^{\prime\prime}$ for the background 
using {\sc ds9} \citep{JoyeDS92003ASPC..295..489J}. Furthermore, we used the {\it grppha} command to group the data with a 
minimum of 200 counts in each bin. We used the data of {\it IXPE} in 2-8 keV and {\it NuSTAR} data in $3 - 79$ keV energy ranges for each respective epoch of observation. For spectral analysis, we used {\sc XSPEC}\footnote{https://heasarc.gsfc.nasa.gov/xanadu/xspec/} \citep{Arnaud1996} version 12.12.0. The $\chi^2$ statistic is used for the goodness of the fit throughout the analysis. The best-fitted spectral plots are further rebinned for visual clarity, it neither affects the model parameter values nor the goodness of the fit.

\begin{table*}
\centering
\caption{\label{table:polparams} Results of model-independent polarimetric analysis performed using \texttt{PCUBE} algorithm.}
\begin{tabular}{cccccc}
\hline
Parameter       & 2-3 keV   & 3-4 keV     & 4-6 keV   & 6-8 keV   & $2-8$ keV  \\ 
\hline
Q/I (\%)        &   $1.90\pm0.41$  &  $1.04\pm0.37$   & $0.31\pm0.50$  &  $1.25\pm1.55$   &  $1.19\pm0.27$\\
U/I (\%)        &   $-0.30\pm0.41$  &  $-4.92\pm0.37$  & $-0.75\pm0.50$ &  $-2.43\pm1.55$  & $-0.61\pm0.27$\\
PD (\%)         &   $1.93\pm0.41$  &  $1.15\pm0.37$   & $0.81\pm0.50$  &  $2.74\pm1.55$   & $1.34\pm0.27$\\
PA ($^{\circ}$) &   $-4.60\pm6.08$ &  $-12.60\pm9.28$ & $-33.54\pm17.88$  & $-31.36\pm16.23$ & $-13.69\pm5.85$\\ \hline
MDP$_{99}$ (\%) &  1.24  &  1.13 &  1.54  &  4.71  &  0.83\\ 
\hline
\end{tabular}
\end{table*}

\section{Results and Discussions}

\subsection{Model-independent {\it IXPE} Analysis}
The \texttt{IXPEOBSSIM} software v30.6.3\footnote{However, we note that a different version of the software gives different results, specifically, a non-detection of polarization using v30.6.4. Therefore, these results are solely based on the above-mentioned version.} \citep{BaldiniEtal2022SoftX..1901194B} is used for the polarimetric analysis of the Level-2 IXPE data obtained from the HEASARC archive. The \texttt{XPSELECT} task is used to extract the source event lists. We did not carry out background subtraction for this source since it is very bright with average flux $\sim$ $1.40 \pm 0.02 \times 10^{-8}$ erg cm$^{-2}$ s$^{-1}$. The model-independent \texttt{PCUBE} algorithm of the \texttt{XPBIN} task \citep{KislatEtal2015APh....68...45K} is used to obtain the polarization cube parameters in the 2–8 keV energy range. We also used the \texttt{PHA1}, \texttt{PHA1Q} and \texttt{PHA1U} algorithms of the \texttt{XPBIN} task to obtain the Stokes I, Q and U spectra of the source. To study any possible energy dependence of these parameters, we grouped the $2-8$ keV data into four energy bands of 2-3, 3-4, 4-6, and 6-8 keV energy ranges and extracted the polarization parameters. We used a similar approach as discussed in \cite{2023MNRAS.521L..74C, MondalEtal2024arXiv240314169M}.

\begin{figure*}
    \centering 
    \includegraphics[width=0.46\textwidth, angle=0]{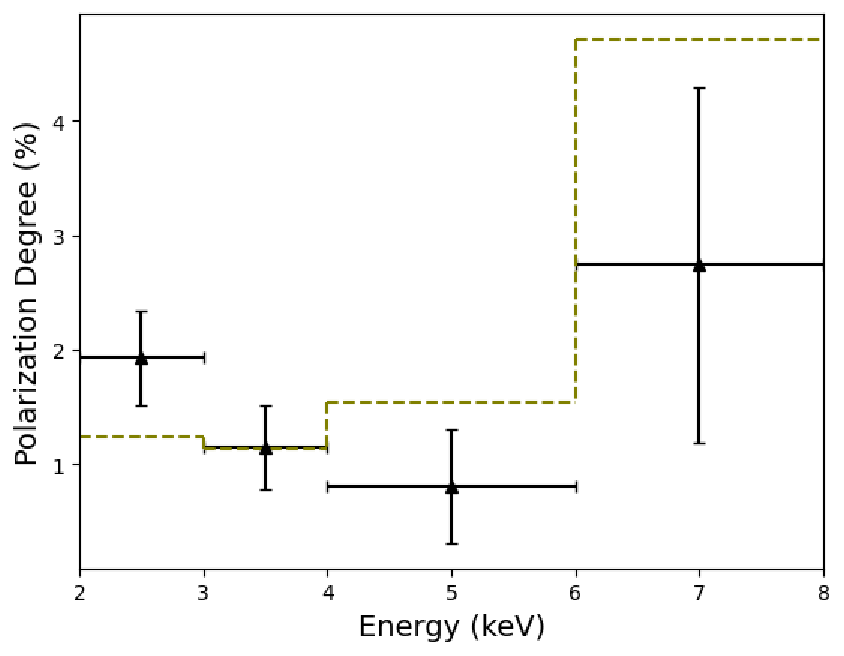}
    \includegraphics[width=0.52\textwidth, angle=0]{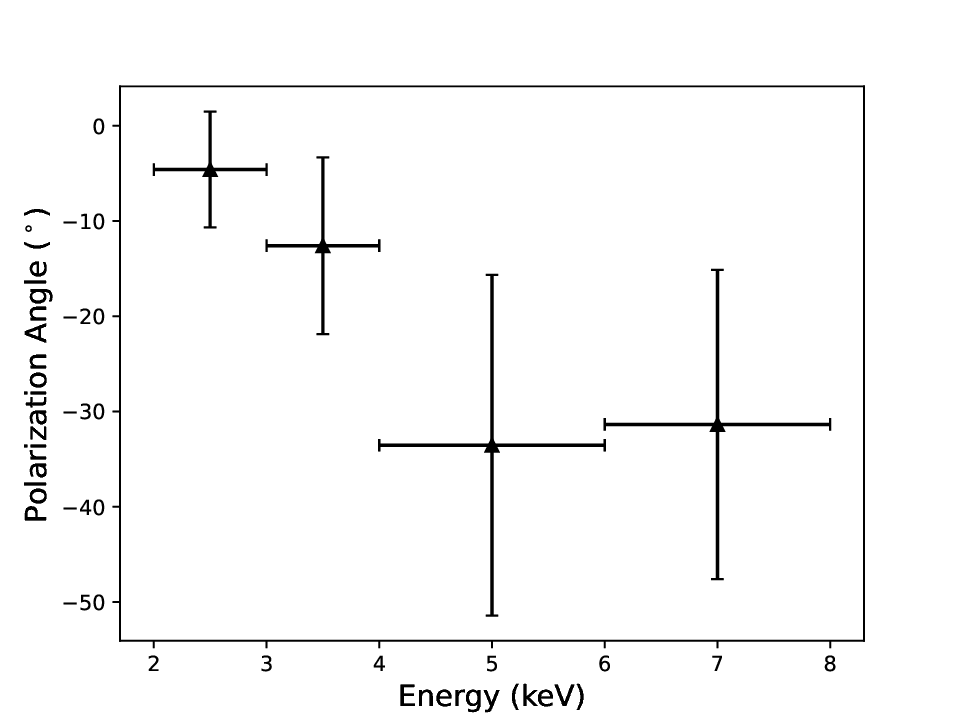}
    \caption{Energy dependent PD and PA plots in $2 - 8$ keV energy range shown with 1 $\sigma$ error bars. The olive dashed line in the left panel represents the estimate of the MDP$_{99}$\%.} 
    \label{fig:pd_and_pa}
\end{figure*}

\begin{figure}
\centering 
\includegraphics[height=8.0truecm,width=9.0truecm]{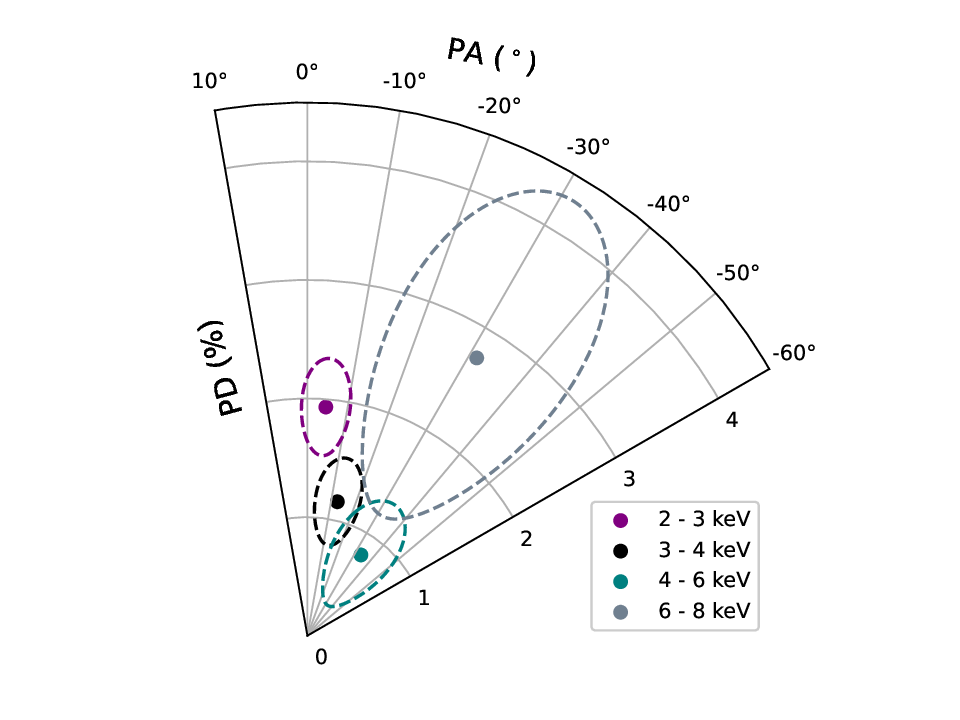};
    \caption{Contour plots between PD and PA obtained using the \texttt{PCUBE} algorithm in 2-3, 3-4, 4-6, and 6-8 keV energy ranges. All contours are drawn with 1 $\sigma$ confidence.} 
    \label{fig:pcube_contours}
\end{figure}

The results of the model-independent polarimetric analysis are summarized in Table \ref{table:polparams}. The measured PD and PA in $2-8$ keV are $1.34\pm0.27$ and $-13.69^\circ\pm5.85^\circ$ respectively. We have also performed energy-dependent PD and PA analyses, which are shown in the left and right panels of Fig. \ref{fig:pd_and_pa} respectively. As it can be observed, the polarization measurements decrease with increasing energy and drop below the MDP$_{99}$ \% above 4 keV. This also gives the PD upper limit (99\% confidence level, shown in olive dashed line) of $<1.54\%$ and $<4.71\%$ in the 4-6 and 6-8 keV bands. The 1 $\sigma$ confidence contours between PA and PD for different energy bands are shown in Fig. \ref{fig:pcube_contours}, obtained using the \texttt{PCUBE} algorithm. The errors obtained through the \texttt{PCUBE} algorithm do not include systematic \citep{KislatEtal2015APh....68...45K}.

Our detected low PD for the source SwiftJ151857 in the SS is in agreement with the measurements of PD from other BHXRBs in the same state, e.g., $\sim 0.5$\% in Swift J1727.8–1613 \citep{SvobodaEtalSwiftJ17272024ApJ...966L..35S} and $\sim 1.95$\% in 4U 1957+115 \citep{Marra2023arXiv231011125M}. Additionally, an energy-dependent PD variation shows a decreasing trend, where the first two energy bands are above or within the MDP, while the higher energy band measurements are below the MDP. Such a decreasing profile in PD has been obtained from theory as well as from numerical simulations and that can be due to the following reasons. The spectra from accretion disks around BHs consist of mainly two types of photons, the low-energy photons that are emitted from the outer region of the disk, where the temperature is lower and the effect of gravity is less important, whereas high-energy emissions are coming from the inner hot region of the disk, where the photon polarization is mostly affected by general relativistic effects. It was shown from numerical simulations that for higher optical depths \citep[see e.g.,][]{DovciakEtal2008MNRAS.391...32D,Krawczynski2012ApJ...754..133K,TavernaEtal2020MNRAS.493.4960T,Marra2023arXiv231011125M} which is the case in the SS, direct radiation turns out to be polarized perpendicularly to the disk symmetry axis at lower energies, while it slowly decreases at higher energies ($>2$ keV) under the effect of the polarization plane rotation. Such turnover energy is decided depending on the spin of the BH and the disk inclination. For higher spinning or inclination systems, turnover occurs at much closer energy and both PA and PD start increasing. However, our source has low inclination and spin (see Sec. 3.4), therefore, the decreasing PA and PD profiles and turnover around 5 keV are well in agreement with the theory and simulation.

Most of the IXPE observations for other BHXRBs showed nearly a constant PA. However, we found a decrease in PA with energy, which is similar to what was observed in 4U 1957+115 \citep{Marra2023arXiv231011125M,KushwahaEtal4U19572023arXiv231002222K}. Such a change in PA could be an indication of rotation of PA between 2-8 keV due to the effects of depolarization. Another possibility can be due to the orientation of jets. However, at present, the jet position angle is not well constrained for this source to compare with the detected PA, requiring a further jet study.

\subsection{Model-dependent {\it IXPE} Analysis}
\begin{figure}
\hspace{-1.5cm}
    \centering 
    \includegraphics[height=9.0truecm,width=10.0truecm]{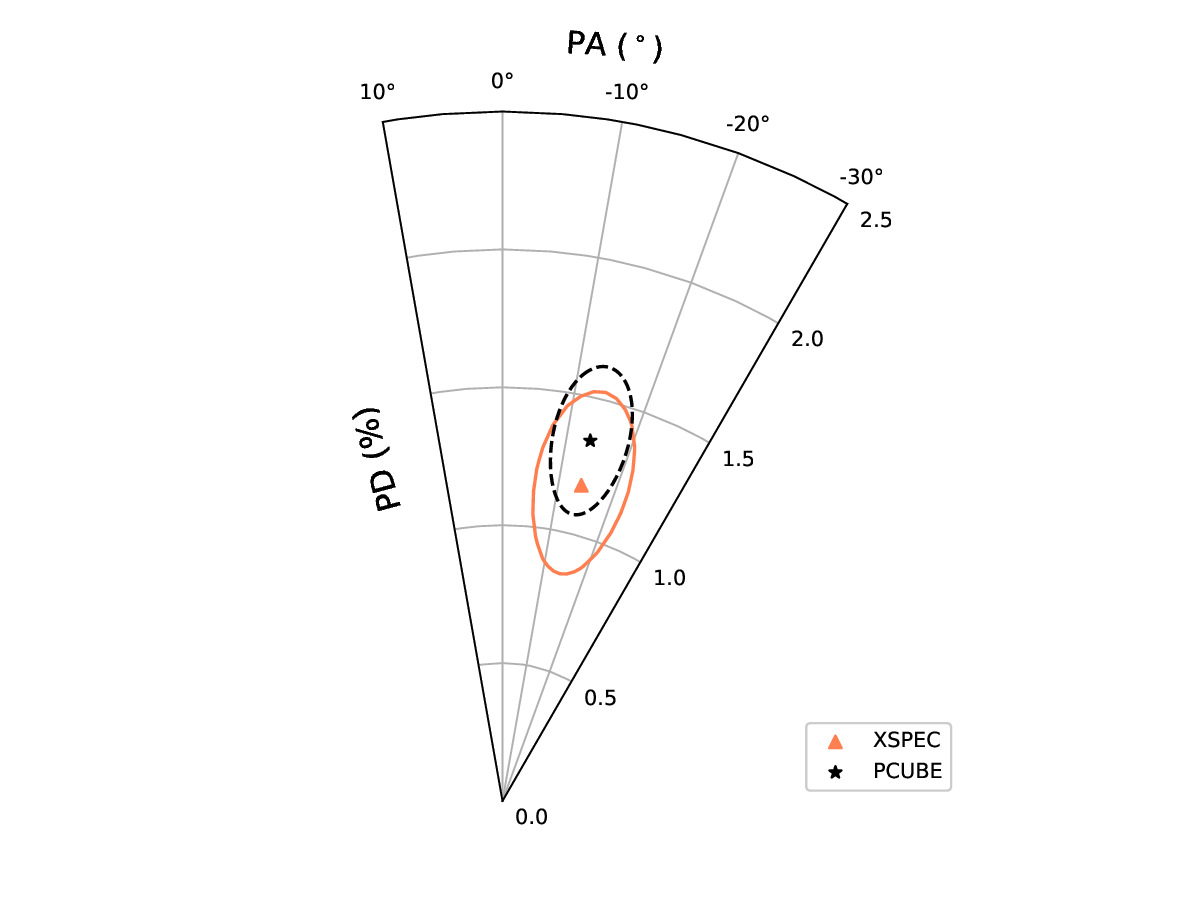};
    \caption{Contour plot between PD and PA with 1 $\sigma$ confidence, obtained using \texttt{PCUBE} algorithm (model-independent) and {\it XSPEC} (model-dependent) fitting method.} 
    \label{fig:xspec_pcube_contours}
\end{figure}

For the model-dependent polarization study, we have fitted all Stokes spectra for the epoch I using absorbed \plaw+\dbb\, model convolved with a constant polarization \polcon\, which reads in {\it XSPEC} as \constant*\tbabs*\polcon*(\plaw+\dbb). All Stokes spectra for all three DUs in the energy band $2-8$ keV are fitted simultaneously. The above model gives an acceptable fit with $\chi^2/dof = 1706/1332$. We determine the PA = $-14.01^\circ\pm5.80^\circ$ and the PD (\%) = $1.18\pm0.23$. Both these values are in excellent agreement with the values obtained from \texttt{PCUBE} algorithm. The neutral hydrogen column density $N_H$ is $5.58\pm0.13\times10^{22}$cm$^{-2}$, obtained from \tbabs\,model.  The \plaw\, model photon index ($\Gamma$) and disk temperature in \dbb\, model are $3.68\pm0.08$ and $0.97\pm0.02$ keV respectively. High values of these two parameters may indicate that the source possibly moved to the soft spectral state during this observation epoch. The source was very recently studied in a follow-up work by \citet{ChatterjeEtal2024arXiv240617629C} using {\it Insight-HXMT} data and observed the disappearance of QPOs in the SS during the epochs that coincide with {\it IXPE} observations, thereby further confirms our finding.  
The confidence contours from the model-dependent analysis are shown in Fig. \ref{fig:xspec_pcube_contours}. The solid and dashed contours drawn using {\it XSPEC} and \texttt{PCUBE} respectively correspond to 1$\sigma$ confidence.

Moreover, as the source moved to the SS, the inner disk moved much closer to the BH, and the corona became highly shrunk, 6-9 $r_S$ ($2GM_{\rm BH}/c^2$, see sec 3.5) in comparison to HS in other BHXRBs \citep[few 100 $r_S$; see][and references therein]{Mondaletal2014,DebnathElal2015MNRAS.447.1984D}. In the SS, it is widely believed that the accretion disk extends down to the innermost stable circular orbit \citep[ISCO;][for a review]{Reynolds2019} and the spectrum is dominated by photons emitted from the very inner region, where the strong gravity can significantly alter the observed polarization properties \citep[][and references therein]{Krawczynski2012ApJ...754..133K,TavernaEtal2020MNRAS.493.4960T}. In this regime, the general relativistic effects cause the photon polarization vectors to undergo rotation, resulting in a net depolarization of the emission at infinity and an overall rotation of the PA. These effects are expected to be stronger for the high-energy photons emitted closer to the central BH, introducing the observed energy dependence of PD for this source in the SS. Numerical simulation performed in the literature \citep{SchnittmanKrolik2010ApJ...712..908S,Krawczynski2012ApJ...754..133K,TavernaEtal2020MNRAS.493.4960T,Marra2023arXiv231011125M}, showed a similar decreasing profile of PD and PA with energy and that also explains the case of low-spinning BH, which is valid for our source. Therefore, we do expect such a profile in PD for our source, if possible to other sources in the future, during {\it IXPE} era. However, for a rapidly spinning BH with high inclination, an opposite PD profile is expected that agrees with the observed PD profile in e.g. 4U 1957+115 in the SS.

We note that polarization measurement changes with spectral states, as we mentioned earlier. Two sources showed a significant change in PD in different spectral states, one is Cyg X-1 \citep{KrawczynskiEtal2022Sci...378..650K,JanaIXPE2024MNRAS.52710837J} and the other Swift J1727.8-1613 \citep{SvobodaEtalSwiftJ17272024ApJ...966L..35S}. These sources also showed significant detection of polarization in the SS, however, PD in the SS was much less compared to the HS. The number of scattering in the SS increases as photons get trapped by the scattering medium, thereby, polarization may decrease. \citet{Podgorny2023MNRAS.526.5964P} detected $\sim1.1$\% polarization in the SS of LMC X-1 and preferred a slab geometry of the corona. Our detection of PD in the SS is in accord with the above findings and further supports that the changes in spectral states are possibly followed by the changes in polarization properties. The possible origin of such low polarization favors the scenario in which the X-ray polarization is driven by the geometry of the accretion flow in the innermost region and an optically thick corona. Also, the polarization of the disk is perpendicular to that of the corona. As the disk emission dominates in the SS, the PD could decrease as was explained in \citet{KrawczynskiEtal2022Sci...378..650K}.

\subsection{Joint {\it IXPE} and {\it NuSTAR} Analysis}
\begin{figure*}
    \centering{
    \hspace{-1.5cm}
    \includegraphics[height=8.0truecm,width=6.0truecm,angle=270]{IN1-kerr-R1.eps}
    \hspace{-0.3cm}
    \includegraphics[height=8.0truecm,width=6.0truecm,angle=270]{IN2-kerr-R1.eps}}
    \caption{Joint spectral modelling of {\it IXPE} and {\it NuSTAR} data fitted using \cutoffpl+\kerrbb\,model in the energy band $2-79$ keV. The left and right panels are for the epochs (I+N1) and (I+N2). The Fe K$\alpha$ line $\sim 6.4$ is fitted using \gauss\, model. The blue, red, and green points correspond to {\it IXPE} data for all three DUs and the black points correspond to {\it NuSTAR} data. Plots are rebinned for visual clarity. See the text for details.
    } 
    \label{fig:SpecFit}
\end{figure*}

\begin{figure*}
    \centering{
    \hspace{-1.5cm}
    \includegraphics[height=8.0truecm,width=6.0truecm,angle=270]{IN1-jetcaf-R1.eps}
    \hspace{-0.3cm}
    \includegraphics[height=8.0truecm,width=6.0truecm,angle=270]{IN2-jetcaf-R1.eps}}
    \caption{Joint spectral modelling of {\it IXPE} and {\it NuSTAR} data fitted using \dbb+\jetcaf\,model in the energy band $2-79$ keV. The left and right panels are for the epochs (I+N1) and (I+N2). The Fe K$\alpha$ line $\sim 6.4$ is fitted using \gauss\, model. The blue, red, and green points correspond to {\it IXPE} data for all three DUs and the black points correspond to {\it NuSTAR} data. Plots are rebinned for visual clarity. See the text for details.
    } 
    \label{fig:SpecJetcaf}
\end{figure*}

\begin{table*}
     \centering
     \caption{Joint {\it IXPE} and {\it NuSTAR}, and independent {\it NuSTAR} spectra of Swift\,J151857 fitted \cutoffpl+\kerrbb\,model parameters are provided in this table. We have fixed the following parameters: distance $D$ = 10 kpc, spectral hardening factor ($hd$) = 1.7, Gaussian Fe K line energy = 6.4 keV. $N_H$ obtained from both fits is $\sim 4-5\times10^{22}$ cm$^{-2}$.}
     \begin{tabular}{cccccccccccc}
     \hline
     Model$\rightarrow$& \multicolumn{4}{c}{\kerrbb} & \multicolumn{2}{c}{\cutoffpl}   &  \gauss \\
     \cline{2-8} 
     Epoch  & $a$ & $i$ & $M_{\rm BH}$ & $\dot m_{\rm d}$ & $\Gamma$  &  $E_{\rm C}$ &  $\sigma_g$ &  $\chi^2/dof$ \\
    $\downarrow$&    & ($^\circ$) & ($M_\odot$) &  ($10^{18}$ gm s$^{-1}$) &    & (keV) &  (keV) &  \\     
      \hline
      I+N1   &  $0.65\pm0.17$ & $35.2\pm7.4$  & $9.2\pm1.6$ & $3.76\pm0.46$ &  $2.47\pm0.06$ & $78.4\pm9.8$ & $0.27\pm0.12$ & 1065/940 \\
      I+N2   & $0.62\pm0.08$ & $38.4\pm6.3$  & $9.7\pm1.3$ &  $4.04\pm0.47$ & $2.37\pm0.10$  & $51.2\pm8.9$ & $0.67\pm0.13$ &872/825 \\
       N1   &  $0.69\pm0.17$ & $46.4\pm15.0$  & $10.1\pm1.7$ & $2.97\pm0.31$ &  $2.30\pm0.08$ & $57.1\pm6.7$ & $0.28\pm0.07$ & 596/504 \\
       N2   & $0.60\pm0.12$ & $40.0\pm9.3$  & $9.3\pm1.4$ &  $4.03\pm0.19$ & $2.19\pm0.03$  & $35.7\pm4.3$ & $0.41\pm0.11$ &421/389 \\
      \hline     
     \end{tabular} 
     \label{tab:spectralfit}
 \end{table*}

For the broadband spectral analysis of the source, we performed a joint fitting using the simultaneous observations from {\it IXPE} in the 2-8 keV range and {\it NuSTAR} in the $3-79$ keV energy range. We used a \constant\, component for a cross-calibration factor between \textit{NuSTAR} and \textit{IXPE}. 
The broadband energy range ($2-79$ keV) covered by the spectra of Swift\,J151857 showed a relatively broad Fe K$\alpha$ line for which a Gaussian component (\gauss\,model) is used at $\sim 6.4$\,keV. We included a \tbabs\, \citep{Wilmsetal2000} component to account for interstellar absorption. The data analysis is also performed using {\it NuSTAR} data in isolation as well. We carried out spectral analysis using both relativistic and non-relativistic accretion disk models independently to estimate the accretion parameters as well as the intrinsic parameters of the central BH.

\subsection{\kerrbb\,Model}
A \kerrbb\,\citep{Li2005ApJS..157..335L} component is used to characterize the emission from an accretion disk, including relativistic effects. This component models the radiation from a geometrically thin and optically thick accretion disk, as proposed by \cite{Novikov&Thorne1973blho.conf..343N}. The \cutoffpl\, component was included to account for the Comptonized emission which originates from the inner hot region of the disk or the corona.  The total model reads in XSPEC as \constant*\tbabs*(\gauss+\cutoffpl+\kerrbb). To take into account the calibration issues, as noted for other black hole systems \citep[see e.g.][]{Marra2023arXiv231011125M,Podgorny2023MNRAS.526.5964P, Rodriguez_Cavero_2023}, we used the \texttt{gain fit} command in \textit{XSPEC} to adjust the response file gains of both telescopes. The same procedures were applied for both the epochs.

In Fig. \ref{fig:SpecFit}, we show the model-fitted spectra of Swift\,J151857 for the joint {\it IXPE} and {\it NuSTAR} observations. The left and right panels show the spectral fits for the epochs I+N1 and I+N2. The lower panels show the residuals. As we did not have the mass, spin parameters of the BH, and the disk inclination beforehand, we left them free during model fitting, keeping the distance to the source fixed at 10 kpc. However, parameter values may change if the source distance changes significantly. Both data fits returned consistent values of all three parameters in both epochs. The mass ($M_{\rm BH}$) and spin ($a$) parameters estimated from the fit vary between $\sim9.2\pm1.6-10.1\pm1.7 M_\odot$ and $0.60\pm0.12-0.69\pm0.17$ respectively. It possibly indicates that the BH is moderately rotating. The disk inclination ($i$) and the mass accretion rate obtained from the fit vary between $\sim 35^\circ\pm7^\circ-46^\circ\pm15^\circ$ and ($3.0\pm0.3-4.0\pm0.5)\times 10^{18}$ gm s$^{-1}$. The low cutoff energy ($E_c$) of $36\pm4-78\pm10$ keV with a photon index ($\Gamma$) of $\sim2.19\pm0.03-2.47\pm0.06$ indicates a transition to the soft spectral state of the source during the observation epoch after ejecting a high radio flare in the first week of March. The $N_H$ value of the source was high $\sim 3.6\pm0.5-5.3\pm0.4\times10^{22}$ cm$^{-2}$, which can be due to strong outflows from the outer disk or due to the presence of some blobs along the line of sight that can block the central radiation \citep[see][]{NeilsenHoman2012ApJ...750...27N,MondalJithesh2023MNRAS.522.2065M}, requires further study of the outflows from the outer disk. We observed a broad Fe K$\alpha$ width which varies between $0.27-0.67$ keV. It is expected as the source moved to the SS, the disk also moved closer to the BH, therefore, relativistic effects can broaden the line \citep[see e.g.][and references therein]{Fabianetal1989,Iwasawaetal1996,BrennemanRey2006ApJ...652.1028B,Mondaletal2016}. How close has the disk moved inward toward the BH? What is the corona geometry (radius and height)? This has been studied in the next section.

\begin{table*}
    \centering
    \caption{Joint {\it IXPE} and {\it NuSTAR}, and independent {\it NuSTAR} spectra fitted \dbb+\jetcaf\,model parameters are provided in this table. All data sets required \gauss\, model for the Fe K$\alpha$ line $\sim 6.4$ keV of width $\sigma_g$ given in the table. $N_H$ obtained from both fits is $\sim 4-5\times10^{22}$ cm$^{-2}$.}
\begin{tabular}{ccccccccccc}
\hline
Model$\rightarrow$&\dbb&\multicolumn{6}{c}{\jetcaf}&\multicolumn{1}{c}{\gauss}\\
\cline{2-9}
Epoch &$T_{\rm in}$&$M_{\rm BH}$ &$\dot m_{\rm d}$ & $\dot m_{\rm h}$ & $X_{\rm s}$ & R &$f_{\rm col}$&$\sigma_g$ &$\chi^2/dof$ \\
	$\downarrow$& (keV) & $(M_\odot)$&$(\dot m_{\rm Edd})$&$(\dot m_{\rm Edd})$&$(r_{\rm S})$& & &(keV) \\
\hline
     I+N1   &$0.96\pm0.01$&$9.8\pm1.5$&$0.74\pm0.04$&$0.25\pm0.01$&$8.7\pm1.5$&$6.67\pm0.55$ &$0.06\pm0.02$&$0.41\pm0.13$&1048/939 \\
     I+N2   &$0.96\pm0.02$&$9.6\pm1.2$&$0.72\pm0.03$&$0.23\pm0.02$&$7.9\pm1.1$&$4.97\pm0.21$ &$0.09\pm0.01$&$0.63\pm0.17$&890/824 \\
     N1   &$0.97\pm0.04$&$9.3\pm1.3$&$0.86\pm0.03$&$0.27\pm0.02$&$6.6\pm1.1$&$5.77\pm0.43$ &$0.06\pm0.01$&$0.27\pm0.09$&583/503 \\
     N2   &$0.94\pm0.03$&$9.2\pm1.0$&$0.81\pm0.04$&$0.23\pm0.02$&$6.0\pm1.1$&$5.48\pm0.37$ &$0.09\pm0.01$&$0.54\pm0.14$&416/388 \\
    \hline     
    \end{tabular} 
    \label{tab:parsJetcaf}
\end{table*}

The \kerrbb\,model fitted parameters infer that the source is possibly a moderately spinning BH with a low disk inclination. It has an impact on getting low polarization (PD) measurements in {\it IXPE} data, as discussed in sec 3.1 and 3.2. If we compare other sources which were observed with high PD in the SS e.g. 4U\,1630-47 ($i \sim 75^\circ$) and LMC\,X-3 ($i \sim 69^\circ$ ) are high-inclination, on the contrary, Cyg\,X-1 and Swift\,J1727.8-1613 showed low PD having low inclination. Our source falls in the second type and the low PD can be attributed either to the repeated scattering of the trapped radiation in an optically thick medium or the polarization of the disk is perpendicular to the corona in the SS \citep[][and references therein]{KrawczynskiEtal2022Sci...378..650K,SvobodaEtalSwiftJ17272024ApJ...966L..35S}.

\subsection{\jetcaf\,Model}
To understand the accretion and outflow properties, we have further fitted the data using an accretion-ejection based two-component advective flow \citep[\tcaf,][]{ChakrabartiTitarchuk1995} model including a jet/outflows component \citep[\jetcaf,][]{MondalChakrabarti2021}. The \jetcaf\, model takes into account the radiation mechanisms at the base of the jet/outflows and the bulk motion effect by the outflowing jet on the emitted spectra, in addition to the Compton scattering of soft disk photons by the hot electron cloud inside the corona. The \jetcaf\,model has six parameters, namely (i) the mass of the BH ($M_{\rm BH}$) if it is unknown, (ii) the Keplerian disk accretion rate ($\dot m_{\rm d}$), (iii) the sub-Keplerian halo accretion rate ($\dot m_{\rm h}$), (iv) the size of the dynamic corona or the location of the shock ($X_{\rm s}$ in $r_S$ unit), (v) the shock compression ratio ($R$), and (vi) the outflow collimation factor ($f_{\rm col}$), the ratio of the solid angle subtended by the outflow to the inflow ($\Theta_o/\Theta_{\rm in}$). The full model reads in {\it XSPEC} for {\it IXPE} and {\it NuSTAR} as \constant*\tbabs*(\dbb+\gauss+\jetcaf).

Fig. \ref{fig:SpecJetcaf} shows the \jetcaf\, model fitted spectra for both epochs. The $\dot m_d$ and $\dot m_h$ values obtained from the fit range between $\sim 0.72\pm0.03-0.86\pm0.03$ and $\sim 0.23\pm0.02-0.27\pm0.02$. The disk moved much closer to the BH with corona radius $\sim 6\pm1-9\pm2$ r$_S$ and $f_{\rm col}$ $\sim 0.06\pm0.01-0.09\pm0.01$. However, the higher value of $\dot m_d$ compared to $\dot m_h$ and lower $X_s$, clearly indicates that the source moved to a soft spectral state. The shock compression ratio ($R$) varies between $5.5\pm0.4-6.7\pm0.6$, and falls in the range where the theoretical mass outflow rate is much less \citep[see][]{Chakrabarti1999}. The mass of the BH ($M_{\rm BH}$) obtained from the fit ranges $9.2\pm1.0$-$9.8\pm1.5$ M$_\odot$ for both epochs, which are constant within the error bar and also agrees with \kerrbb\,model fitted BH mass. All model fitted parameters are given in Table \ref{tab:parsJetcaf}.

Given the \jetcaf\,model fitted parameters, we have estimated the mass outflow rate for this source to be $0.011\pm0.001$ (minimum for I+N1 parameters) and $0.022\pm 0.003$ (maximum for N2 parameters) $\dot M_{\rm Edd}$ using Eq. 16 derived in \citet[][]{Chakrabarti1999}. In our model, as discussed earlier, both the corona and the base of the jet behave as a Comptonizing medium. Such a low mass outflow rate also agrees with the soft spectral state. 

The \jetcaf\,model fitted hot accretion flow component ($\dot m_h$) that contributes to the power law emission from the corona is $\sim 1/3$ of the cold thermal disk component ($\dot m_d$). Additionally, the size of the corona ($X_s$) is significantly small ($6-9 ~r_S$), compared to its size in the intermediate states which is $\gtrsim 60~r_S$, phenomenologically estimated using timing (QPO) properties \citep{ChatterjeEtal2024arXiv240617629C}. These estimates infer that the emission is dominated by the thermal disk rather than the corona. Therefore, the geometrical configuration and high disk mass accretion rate may explain that the observed low PD mainly comes from the accretion disk, where the polarization is perpendicular to the disk. Conversely, the PD contribution from the corona is expected to be less due to its smaller size loaded with higher mass, which can trap radiation by repeated scattering resulting in lower PD.

Subsequently, the unpolarized radiation from the central source scattered by the corona of radius $X_s$ and height $h_{\rm shk}$ can gain a net fractional polarization, depending on $i$. Therefore, given the best-fitted corona geometry and $i$, one can estimate the PD using the relation \citep{BegelmaMcKee1983ApJ...271...89B}, $PD = \sin^2i (1-\mathcal{R_{\rm c}})/[2(1+\mathcal{R_{\rm c}})+(1-\mathcal{R_{\rm c}})\sin^2i]$, where $\mathcal{R_{\rm c}}=2(h_{\rm shk}/X_{s})^2$. The $h_{\rm shk}$ is estimated using Eq. 1 derived in \citet[][]{Debnathetal2014}. For the best-fitted model parameters in epoch I+N2, the above relation predicts expected PD $\sim 4-6\%$. This range of PD was detected for other sources earlier (see Sec. 1) in their hard or intermediate spectral states, where the corona properties change and that can contribute more to polarization. Therefore, such predictions can be further verified in the future if the source can be observed in other spectral states independently by {\it IXPE} and India's recently launched X-ray Polarimeter Satellite {\it XPoSat}, or jointly.

\section{Conclusions}
In this paper, we have performed spectro-polarimetric studies of a recently discovered black hole X-ray binary SwiftJ151857 using simultaneous {\it IXPE} and {\it NuSTAR} observations during 17 and 18 March 2024. The key findings from this study are summarized below:

\begin{itemize}
   \item The first detection of X-ray polarization of the source using {\it IXPE} yields PD and PA $\sim1.34\pm0.27\%$ and $\sim -14.0^\circ\pm5.8^\circ$ in the soft spectral state. The PD value further supports that the changes in spectral states are possibly followed by changes in polarization properties.

   \item The low PD measurement can be due to the low spin and inclination of the disk which agrees with the previous detections. Additionally, in the soft state, the corona has significantly shrunk increasing its optical depth and the emission is coming mainly from the disk. Therefore, it can significantly depolarize the emission by repeated scattering.

   \item The low cutoff powerlaw energy ($\sim 36-78$ keV) and high powerlaw photon index ($>2.2$) indicate a transition to a soft spectral state of the source. This is expected as the source ejected a high radio flare in the first week of March and later moved to the soft spectral state. The spectral state has further been confirmed from the non-detection of QPOs using {\it Insight-HXMT} observations during this epoch \citep{ChatterjeEtal2024arXiv240617629C}.

    \item Additionally, the high disk mass accretion rate compared to the hot flow rate with lower corona size ($\sim 6-9$ $r_S$) supports the transition of the source to the soft spectral state.

    \item The mass of the black hole estimated from different model fits in a narrow range between $9.2\pm1.0-10.1\pm1.7$ M$_\odot$. The spin parameter of the source is $\sim 0.6\pm0.1-0.7\pm0.2$ with accretion disk inclination $\sim36^\circ\pm7^\circ-46^\circ\pm15^\circ$ for a source distance of 10 kpc.

    \item We did not find a significant mass outflow rate ($<0.03$ $\dot M_{\rm Edd}$), which can be due to the transition of the source to the soft spectral state.
\end{itemize}

We have detected X-ray polarization of the source SwiftJ151857 for the first time in the soft spectral state and inferred the origin of polarization properties (PD and PA profiles). Our detailed study of the BHXRB source SwiftJ151857 sheds light on the accretion disk properties and the geometry of the corona, as well as the estimation of intrinsic properties of the black hole. The estimated spin parameter and disk inclination can explain the low PD detection. Moreover, the joint {\it IXPE} and {\it NuSTAR}, and independent {\it NuSTAR} analysis show consistent results. A detailed timing study on this source further confirms the spectral state of the source \citep{ChatterjeEtal2024arXiv240617629C}. We have further predicted the PD using both model-fitted parameters, which is significantly high (4-6\%), however, in the range of PD detected for other sources in their hard or intermediate states. Such predictions can be further verified in the future using {\it IXPE} and {\it XPoSat}, independently or jointly.  

\section*{Acknowledgements}
We thank the referee for making fundamental queries on the physical understanding of the results, which improved the quality of the paper.
SM and SPS acknowledge Ramanujan Fellowship (\# RJF/2020/000113) by SERB-DST, Govt. of India for this research. K.C. acknowledges support from SWIFAR postdoctoral fund of Yunnan University. CBS is supported by the National Natural Science Foundation of China under grant no. 12073021. This research used data products provided by the IXPE Team
(MSFC, SSDC, INAF, and INFN) and distributed with additional
software tools by the High-Energy Astrophysics Science Archive Research Center (HEASARC), at NASA Goddard Space Flight Center (GSFC). This research has made use of the {\it NuSTAR} Data Analysis Software (NuSTARDAS) jointly developed by the ASI Science
Data Center (ASDC, Italy) and the California Institute of Technology (Caltech, USA).


\bibliographystyle{aasjournal} 
\bibliography{example}






\end{document}